\documentclass[preprint,aps,prd,nofootinbib,showpacs,superscriptaddress,11pt]{revtex4}
\usepackage{txfonts}
\usepackage{float}
\usepackage{graphicx}
\usepackage{dcolumn}
\usepackage{bm}
\usepackage{amssymb}
\usepackage{latexsym}

\newcommand{\be}{\begin{equation}}
\newcommand{\ee}{\end{equation}}
\newcommand{\bq}{\begin{eqnarray}}
\newcommand{\eq}{\end{eqnarray}}

\begin{document}

\preprint{USTC-ICTS-12-04}

\title{Generalized Holographic Dark Energy and its Observational Constraints}

\author{Zhenhui Zhang}
\email{zhangzhh@mail.ustc.edu.cn} \affiliation{Department of Modern
Physics, University of Science and Technology of China, Hefei
230026, China} \affiliation{Institute of Theoretical
Physics, Chinese Academy of Science, Beijing 100190, China}
\author{Miao Li}
\email{mli@itp.ac.cn} \affiliation{Institute of Theoretical Physics,
Chinese Academy of Science, Beijing 100190, China}
\affiliation{Kavli Institute for Theoretical Physics China, Chinese
Academy of Sciences, Beijing 100190, China}
\affiliation{State Key Laboratory of Frontiers in Theoretical Physics, Chinese Academy of Sciences, Beijing 100190, China}
\author{Xiao-Dong Li}
\email{renzhe@mail.ustc.edu.cn} \affiliation{Department of Modern
Physics, University of Science and Technology of China, Hefei
230026, China}  \affiliation{Institute of Theoretical
Physics, Chinese Academy of Science, Beijing 100190, China}
\affiliation{Interdisciplinary
Center for Theoretical Study, University of Science and Technology
of China, Hefei 230026, China}
\author{Shuang Wang}
\email{swang@mail.ustc.edu.cn} \affiliation{Department of Modern
Physics, University of Science and Technology of China, Hefei
230026, China} \affiliation{Institute of Theoretical Physics,
Chinese Academy of Science, Beijing 100190, China}
\author{Wen-Shuai Zhang}
\email{wszhang@mail.ustc.edu.cn} \affiliation{Department of Modern
Physics, University of Science and Technology of China, Hefei
230026, China} \affiliation{Institute of Theoretical Physics,
Chinese Academy of Science, Beijing 100190, China}

\begin{abstract}
In the original holographic dark energy (HDE) model, the dark energy density is proposed to be $\rho_{de} = 3c^2M^2_{pl}L^{-2}$,
with $c$ is a dimensionless constant characterizing the properties of the HDE. In this work, we propose the generalized holographic
dark energy (GHDE) model by considering the parameter $c$ as a redshift-dependent function $c(z)$. We derive all the physical quantities
of the GHDE model analytically, and fit the $c(z)$ by trying four kinds of parametrizations. The cosmological constraints of the $c(z)$
are obtained from the joint analysis of the present SNLS3+BAO+CMB+$H_0$ data. We find that, compared with the original HDE model, the
GHDE models can provide a better fit to the data. For example, the GHDE model with JBP-type $c(z)$ can reduce the $\chi^2_{min}$ of
the HDE model by 2.16. We also find that, unlike the original HDE model with a phantom-like behavior in the future, the GHDE models
can present many more different possibilities, i.e., it allows the GHDE in the future to be either quintessence like, cosmological constant like, or phantom like, depending on the forms of $c(z)$.

\end{abstract}


\maketitle

\section{Introduction}\label{sec:intro}

Since its discovery in 1998 \cite{Riess}, dark energy (DE)
\cite{DEReview} has become one of the most popular research areas in
modern cosmology. Numerous theoretical models have been proposed in
the last decade
\cite{quint,phantom,k,Chaplygin,tachyonic,hessence,YMC,Onemli}.
However, the nature of DE still remains a mystery.

A popular and interesting approach to the nature of DE is considering it as an issue of quantum gravity \cite{Witten:2000zk}.
Since there is no complete theory of quantum gravity now,
we usually consider  some effective theories,
in which some fundamental principles are taken into account, such as the commonly believed holographic principle \cite{Holography}.
In 1999, based on the effective quantum field theory,
Cohen et al. \cite{cohen99} pointed out that the quantum zero-point energy of a system with size $L$ should not exceed the mass of a black hole with the same size, i.e.,
\begin{equation}
L^3\rho_{\Lambda}\leq L M_{pl}^2~,
\end{equation}
here $L$ is the ultraviolet (UV) cutoff,
which is closely related to the quantum zero-point energy density,
and $M_{pl}\equiv1/\sqrt{8\pi G}$ is the reduced Planck mass.
In this way, the UV cutoff of a system is related to its infrared (IR) cutoff.
It means that the vacuum energy related to this holographic principle may be viewed as DE when we take the whole universe into account (its energy density is denoted as $\rho_{de}$ hereafter).
The largest IR cutoff $L$ is chosen by saturating the inequality, so that we get the holographic dark energy (HDE) density
\begin{equation}\label{eq:rhoHDE}
\rho_{de} = 3c^2M^2_{pl}L^{-2}~,
\end{equation}
where $c$ is a dimensionless model parameter.
If we take $L$ as the size of the current universe,
for instance the Hubble radius $H^{-1}$,
then the DE density will be close to the observational result.
However, Hsu \cite{hsu04} pointed out that this choice yields a wrong equation of state (EOS) for DE.
For this reason,
Li \cite{li04} suggested to choose the future event horizon of the universe as the IR cutoff,
defined as,
\begin{equation}\label{eq:rh}
R_h=a\int^\infty_t{dt \over a}=a\int^\infty_a{da'\over Ha'^2}~.
\end{equation}
This choice  gives a reasonable value for the DE density, and also leads to an accelerated universe.
The HDE model based on this ansatz has been proved to be a competitive and promising DE candidate.
It can theoretically explain the coincidence problem \cite{li04},
and is proven to be perturbational stable \cite{HDEstable}
(see \cite{casimirHDE,Hogan,jwlee,liwang,hologas}
for more theoretical studies).  It is also found that this original HDE model is much more favored by the observational data \cite{holoobs09}, even compared with other holographic dark energy models with different IR cutoffs \cite{refADE}\cite{refRDE}( see \cite{refHDE,holode1,holode2,holodeobs,intholo,healworld} for more researches).

The parameter $c$ in Eq. (\ref{eq:rhoHDE}) plays an essential role in characterizing the dark energy properties in the HDE model,
 e.g. with the value of $c$ being bigger or smaller than 1, the behavior of HDE in the far future would be phantom-like or quintessence-like,  which means giving very different ultimate fates for the universe (see Sec. V for  detail).
In the previous works,  $c$ is always treated as a constant.
However, since there are no confirmed evidence at present telling us that $c$ shall be a constant,
it is worthwhile, and somehow natural, to treat this parameter as a variable (a similar idea has been applied on the holographic ricci dark energy in \cite{Weihao}).
Motivated by this idea, in this paper, we generalize the original HDE model by treating the parameter $c$ as a free function of redshift $z$, i.e.,
\begin{equation}\label{eq:rhoGHDE}
\rho_{de} = 3c(z)^2M^2_{pl}R_h^{-2}.
\end{equation}
Hereafter, we will denote it as ``GHDE'' (generalized holographic dark energy) model.
Not having a theory to fix the form of $c(z)$ at present,it is helpful to take some trial functions to characterize it.
As a test, we consider four parametrizations of $c(z)$ here, i.e.,
\begin{eqnarray}\label{eq:czform}
1.\ {\rm GHDE1}:\ \  c(z)&=&c_0+c_1\frac{z}{1+z}\label{eq:GHDE1c}~;\\
2.\ {\rm GHDE2}:\ \ c(z)&=&c_0+c_1\frac{z}{(1+z)^2}\label{eq:GHDE2c}~; \\
3.\ {\rm GHDE3}:\ \ c(z)&=&\frac{c_0}{1+c_1\ln(1+z)}\label{eq:GHDE3c}~; \\
4.\ {\rm GHDE4}:\ \  c(z)&=&c_0+c_1\left(\frac{\ln(2+z)}{1+z}-\ln2\right)\label{eq:GHDE4c}~.
\end{eqnarray}
 These choices are inspired by those parametrizations proposed to study the EoS of DE,  known as Chevallier-Polarski-Linder parametrization (CPL) \cite{CPL}, Jassal-Bagla-Padmanabhan (JBP) parametrization \cite{JBP}, Wetterich parametrization \cite{Wetterich},
and Ma-Zhang parametrization \cite{zhangxinlog}, respectively. The original HDE model can be recovered when we take $c_1 = 0$ in all these four parametrizations.

In this work, we study the observational constraints for these GHDE models, by fitting the cosmological data from the recently released SNLS3 sample of 472 type Ia supernovae \cite{SNLS3},
the cosmic microwave background anisotropy data from the Wilkinson Microwave Anisotropy Probe 7-yr observations \cite{WMAP7},
the baryon acoustic oscillation results from the Sloan Digital Sky Survey data release 7 \cite{SDSSDR7},
and the Hubble constant measurement from the Wide Field Camera 3 on the Hubble Space Telescope \cite{HSTWFC3}. The original HDE model is also investigated for a comparison.

This paper is organized as follows.
In Sec. II, we derive the basic equations for the GHDE model.
In Sec. III, by fitting data, we investigate the cosmological constraints on four GHDE models each with one form of  $c(z)$ listed in Eq. (\ref{eq:czform}).
In Sec. IV, we discuss the different fates of the universe in both the HDE and GHDE models.
At last, we give some concluding remarks in Sec. V.
The methodology used in this work is listed in Appendix.

\section{Generalized Holographic Dark Energy Model}\label{GHDE}

In this section, we derive the basic equations for the GHDE model.
We will begin with the basic FRW cosmology, and then introduce the GHDE model subsequently.

\subsection{The FRW cosmology}\label{FRW}

In a spatially flat isotropic and homogeneous Friedmann-Robertson-Walker (FRW) universe
(the assumption of flatness is motivated by the inflation scenario),
the Friedmann equations read
\begin{eqnarray}
3M^2_{pl}H^2 &=& \rho~, \label{eq:FRWEq1} \\
-2M^2_{pl}\dot H &=& \rho + p~,\label{eq:FRWEq2}
\end{eqnarray}
where $H=\dot a/a$ is the Hubble parameter, and $\rho$ is the total energy density of the universe,
including the components of matter $\rho_m$, radiation $\rho_r$ and dark energy $\rho_{de}$.
For simplicity, we will not consider the case with interaction between dark matter and DE in this work.
So we have energy conservation equation for each component,
\begin{equation}\label{eq:matterconserve}
\dot{\rho}_{m}+3H\rho_{m}=0~,
\end{equation}
\begin{equation}\label{eq:radationconserve}
\dot{\rho}_{r}+4H\rho_{r}=0~,
\end{equation}
\begin{equation}\label{eq:deconserve}
\dot{\rho}_{de}+3H(1+w_{de})\rho_{de}=0~.
\end{equation}

Combining with Eqs.(\ref{eq:matterconserve}) and (\ref{eq:radationconserve}), the first Friedmann equation Eq. (\ref{eq:FRWEq1}) can be rewritten as,
\begin{equation}\label{eq:Ez}
E(z)\equiv \frac{H(z)}{H_0}=\left(\frac{\Omega_{m0}(1+z)^3+\Omega_{r0}(1+z)^4}{1-\Omega_{de}(z)}\right)^{1/2}~.
\end{equation}
Here $H_{0}$ is the Hubble constant,
$\Omega_{de}(z)$ is the fractional dark energy density, given by,
\begin{equation}\label{eq:Omegade}
\Omega_{de}(z)\equiv\frac{\rho_{de}(z)}{\rho_c(z)}=\frac{\rho_{de}(z)}{3M^2_{pl}H^2}~,
\end{equation}
and $\Omega_{m0}$ and $\Omega_{r0}$ are the present values of fractional density for matter and radiation,
respectively.
From the WMAP observations, $\Omega_{r0}$ is \cite{WMAP7},
\begin{equation}\label{eq:Omegar}
\Omega_{r0}=\Omega_{\gamma0}(1+0.2271N_{eff})~,\ \ \
\Omega_{\gamma0}=2.469\times10^{-5}h^{-2},\ \ \ N_{eff}=3.04~,
\end{equation}
where $\Omega_{\gamma0}$ is the present fractional photon density,
$h$ is the reduced Hubble parameter,
and $N_{eff}$ is the effective number of neutrino species.

\subsection{The evolution of GHDE}\label{sec:GHDE}

Now let us introduce the HDE scenario into the cosmology.
Since the original HDE model with constant $c$ is a subset of the GHDE model,
we can start straightforwardly with the GHDE model.

First of all, let us derive the effective DE EoS of the GHDE model.
Taking derivative for Eq. (\ref{eq:rhoGHDE}) with respect to $x=\ln a,~a=\frac{1}{1+z}$, and making use of Eq. (\ref{eq:rh}), we get
\begin{equation}\label{eq:drhoGHDEdz}
\rho_{de}^{\prime}\equiv \frac{d\rho _{de}}{dx}
= 2 \rho_{de}\left(\frac{\sqrt{\Omega_{de}}}{c}-1+\frac{1}{c}\frac{dc}{dx}\right)~.
\end{equation}
Combining Eqs.~(\ref{eq:deconserve}) and (\ref{eq:drhoGHDEdz}), it follows the DE EoS,
\begin{equation}\label{eq:GHDEwde}
w_{de}=-{1\over 3}-{2\over 3c}\sqrt{\Omega_{de}}+{2\over 3}\frac{(1+z)}{c}\frac{dc}{dz}~.
\end{equation}
Compared with the original HDE model,
there is an additional term ${2\over 3}\frac{(1+z)}{c}\frac{dc}{dz}$ due to the redshift dependent of $c(z)$.
This additional term means that not only the value, but also the differential of $c(z)$ plays an important role in the evolution of DE. In the next sections, we will find this additional term make GHDE model obviously different with the original HDE model.

Now let us have a look at the property of the GHDE fraction $\Omega_{de}(z)$,
which influences the expansion history of the universe.
To do this, we shall solve out the unknown $\Omega_{de}(z)$ in Eq. (\ref{eq:Ez}).
Directly taking derivative for $\Omega_{de}={c^2/ (H^2L^2)}$, and using Eq. (\ref{eq:rh}), we get
\begin{equation}\label{eq:GHDEdOmegadedx}
\Omega_{de}'=2\Omega_{de}\left(\epsilon-1+{\sqrt{\Omega_{de}}\over c}+\frac{1}{c}\frac{dc}{dx}\right)~,
\end{equation}
where $\epsilon\equiv -{\dot{H}/ H^2}=-{H'/ H}$.
From Eqs.~(\ref{eq:FRWEq2}) and (\ref{eq:GHDEwde}), we have
\begin{equation}\label{eq:epsion}
\epsilon={3\over 2}(1+w_{de}\Omega_{de})
={3\over2}-\left({1\over2}+\frac{1}{c}\frac{dc}{dx}\right)\Omega_{de}-{\Omega_{de}^{3/2}\over c}~.
\end{equation}
Thus, with $\frac{d}{dx}=-(1+z)\frac{d}{dz}$, we have the equation of motion, a differential equation, for $\Omega_{de}$,
\begin{equation}\label{eq:dOmegadedz}
\frac{d\Omega_{de}}{dz}
=-\frac{1}{1+z}\Omega_{de}(1-\Omega_{de})
\left(1+{2\sqrt{\Omega_{de}}\over c}-\frac{2(1+z)}{c}\frac{dc}{dz}\right)~.
\end{equation}
Also, we see an additional term $-\frac{2(1+z)}{c}\frac{dc}{dz}$ in the bracket
due to the redshift-dependent $c$ in the GHDE model.
In the data analysis procedure, this useful Eq.~(\ref{eq:dOmegadedz}) can be solved numerically.
And once we have $\Omega_{de}$, the Hubble parameter can be obtained directly through Eq. (\ref{eq:Ez}).

\section{Observational Constraints on the GHDE Models}\label{fitGHDE}

In this section,
we will place the cosmological constraints on the GHDE models discussed above.
Hereafter we consider four kinds of parametrizitions of $c(z)$ listed in Eq. (\ref{eq:czform}) respectively.

\subsection{GHDE1: the CPL type}

For this case, $c(z)$ takes the form
\begin{equation}\label{eq:GHDE1c}
c(z)=c_0+c_1\frac{z}{1+z}~.
\end{equation}
To be more accurate, this parametrization satisfies: $z\rightarrow 0,c\rightarrow c_0; ~z\rightarrow +\infty,c\rightarrow c_0+c_1.$ i.e.  $c$ smoothly varied from $c_0+c_1$ to $c_0$ from the past to present.
If we assume that the GHDE has positive energy density ($\rho_{de}>0$),
we can obtain the following constraints on the parameters,
\begin{equation}\label{eq:GHDE1_condition}
c_0>0~,\ \ c_0+c_1>0~.
\end{equation}

Taking derivatives for Eq. (\ref{eq:GHDE1c}), it follows that
\begin{equation}
\frac{(1+z)}{c}\frac{dc}{dz}=\frac{c_1}{c_0+(c_0+c_1)z}~.
\end{equation}
Substituting this equation to Eq. (\ref{eq:GHDEwde}) and Eq. (\ref{eq:dOmegadedz}),
one can get the EOS and Friedmann equations in the GHDE1 model.

\begin{figure}
\makeatletter
          \def\@captype{figure}
          \makeatother
\begin{center}
\includegraphics[width=7.4cm]{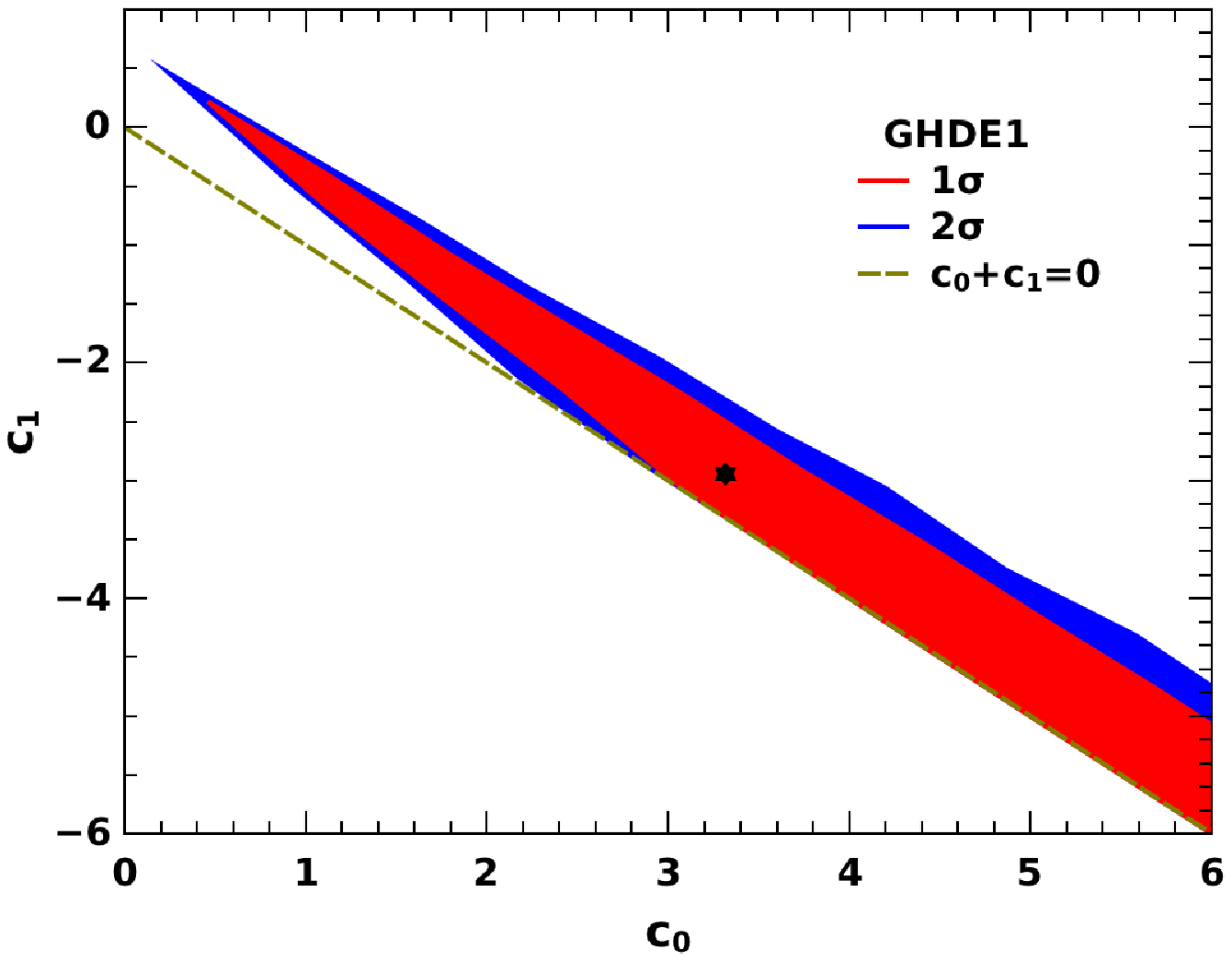}
\includegraphics[width=7.7cm]{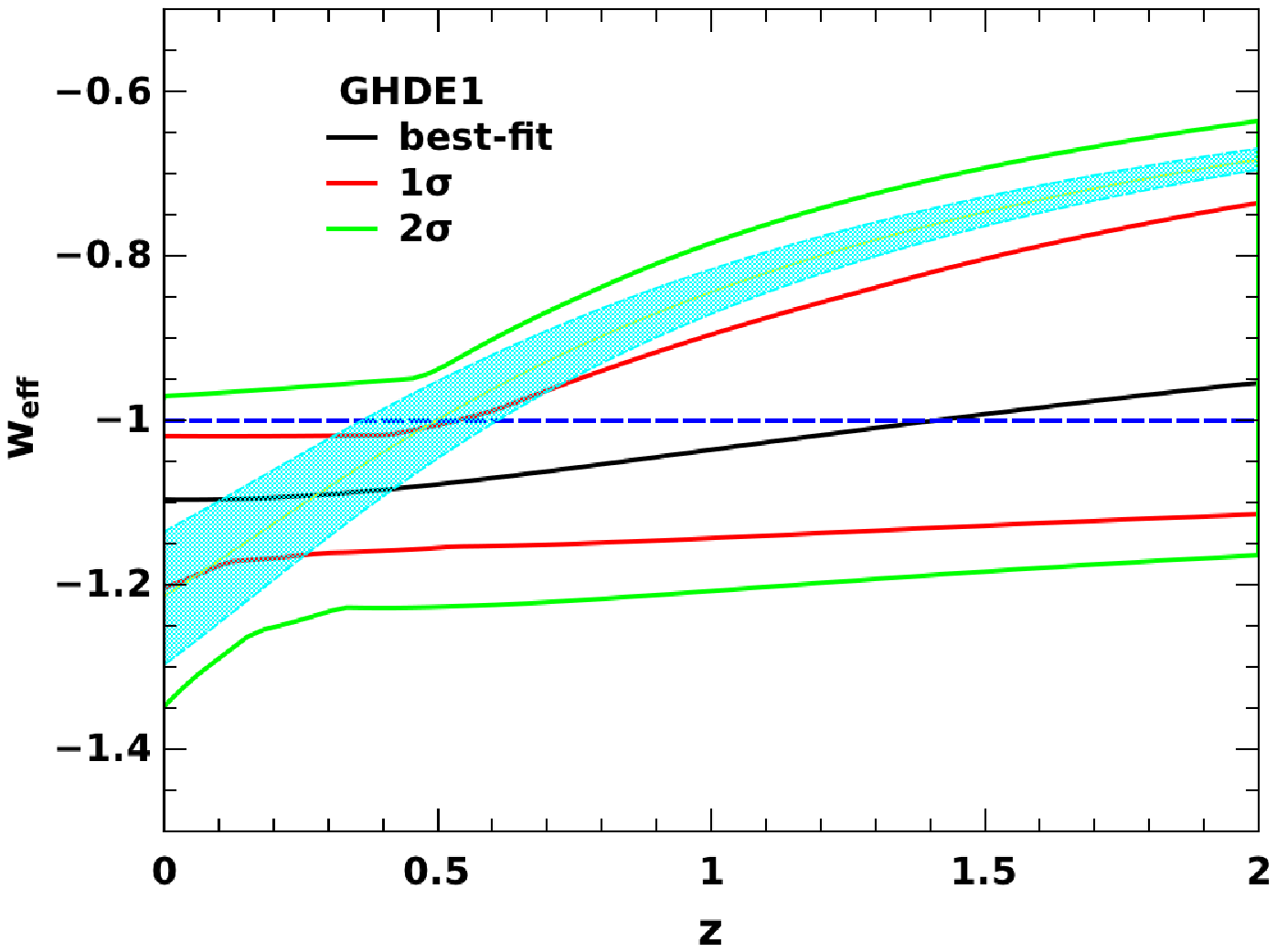}
\end{center}
\caption{\label{figCPL}
$Left\ Panel$: Marginalized probability contours at 1$\sigma$ and 2$\sigma$ CL in the $c_0-c_1$ plane, for the GHDE1 model.
The constraints Eq. (\ref{eq:GHDE1_condition}) are automatically satisfied.
$Right\ Panel$: The evolution of $w_{eff}$ along with $z$, for the same model.
The $w=-1$ line and the 1$\sigma$ constraint of the original HDE model are also shown.
They all lie in the 2$\sigma$ region of the GHDE1 model. }
\end{figure}

In the left panel of Fig.~\ref{figCPL}, we plot the 1$\sigma$ and 2$\sigma$ contours for the GHDE1 model in the $c_0-c_1$ plane.
It is noticeable that the constraints of Eq. (\ref{eq:GHDE1_condition}) (plotted in the black dashed line) is automatically satisfied by the results of the numerical simulation.
Another feature of this model is that there is a strong degeneracy between the parameter $c_0$ and $c_1$.
The constrained parameter space distributes roughly along the line $c_0+c_1=0$.

In the right panel of Fig.~\ref{figCPL}, we show the evolution of $w_{eff}(z)$ of this model( the $w_{eff}$ of the original HDE model is also showed).
Unlike the original HDE model which significant deviates from the $\Lambda$CDM model at $z\lesssim0.3$,
here we find that the $w=-1$ line lies in roughly the 1$\sigma$ error of the evolution of $w_{eff}(z)$,
with only small deviations in the low redshift region.
We also find that the 1$\sigma$ constraints of the $\Lambda$CDM and HDE models all lie in the 2$\sigma$ error of the $w_{eff}$,
so they are all consistent with the GHDE1 model in the 2$\sigma$ CL.

\subsection{GHDE2: the JBP type }
The ``JBP-typ'' GHDE model has the following form of $c(z)$:
\begin{equation}\label{eq:GHDE2c}
c(z)=c_0+c_1\frac{z}{(1+z)^{2}}~.
\end{equation}
And it follows directly that,
\begin{equation}
\frac{(1+z)}{c}\frac{dc}{dz}=\frac{c_1(1-z)}{c_0(1+z)^2+c_1z}~.
\end{equation}

From Eq.~(\ref{eq:GHDE2c}), we get: $z\rightarrow 0, c\rightarrow c_0;~z\rightarrow 1,c\rightarrow c_0+\frac{1}{4}c_1;~z\rightarrow +\infty, c\rightarrow c_0.~$
Like the CPL-type parametrization,
the JBP parametrization is also unable to describe the behavior of HDE in the future.

In this model, the constraint $\rho_{de}>0$ gives,
\begin{equation}
\label{eq:GHDE2_condition}
c_0> 0~,\ \ \ c_0+\frac{1}{4}c_1>0~.
\end{equation}

\begin{figure}
\begin{center}
\includegraphics[width=7.35cm]{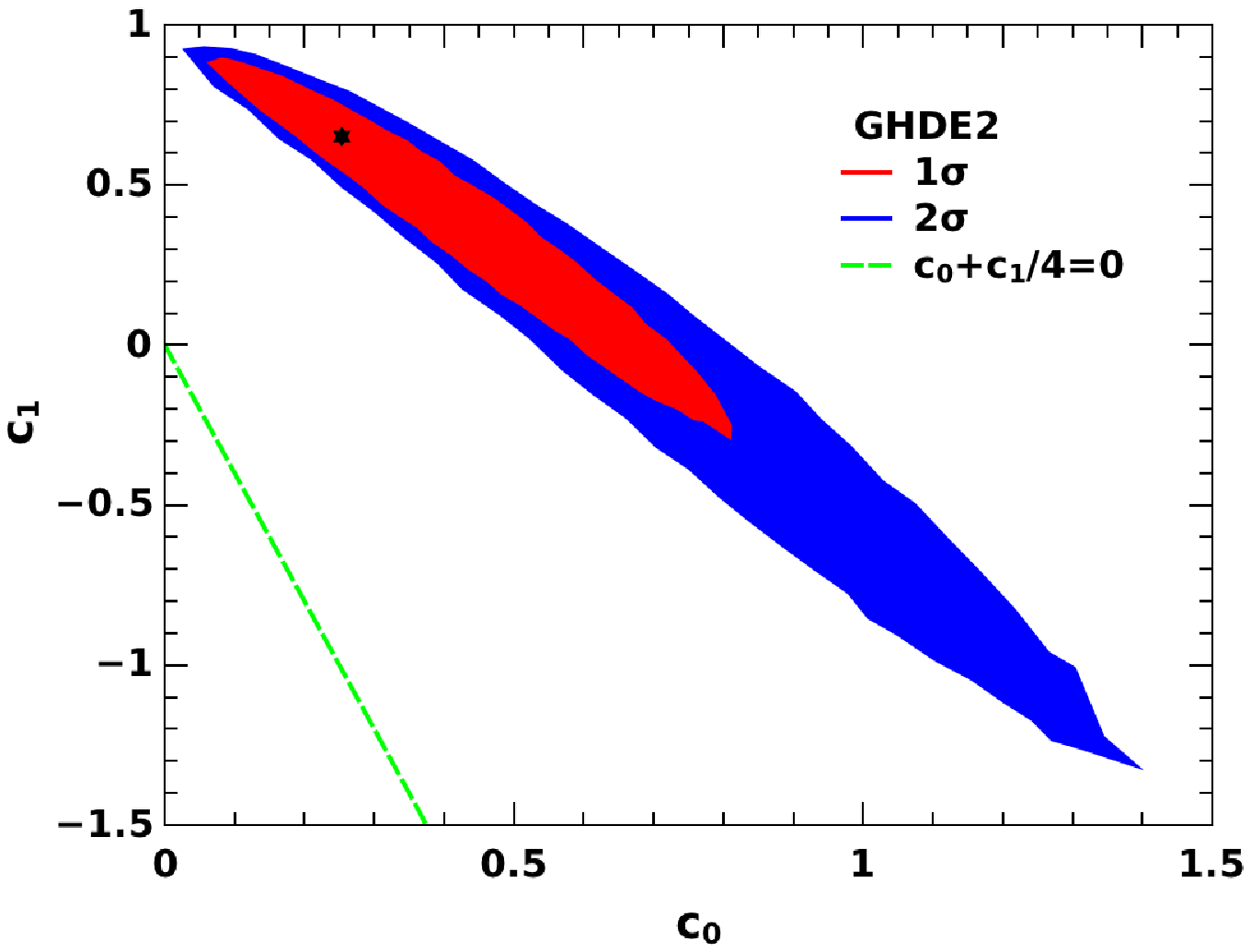}
\includegraphics[width=7cm]{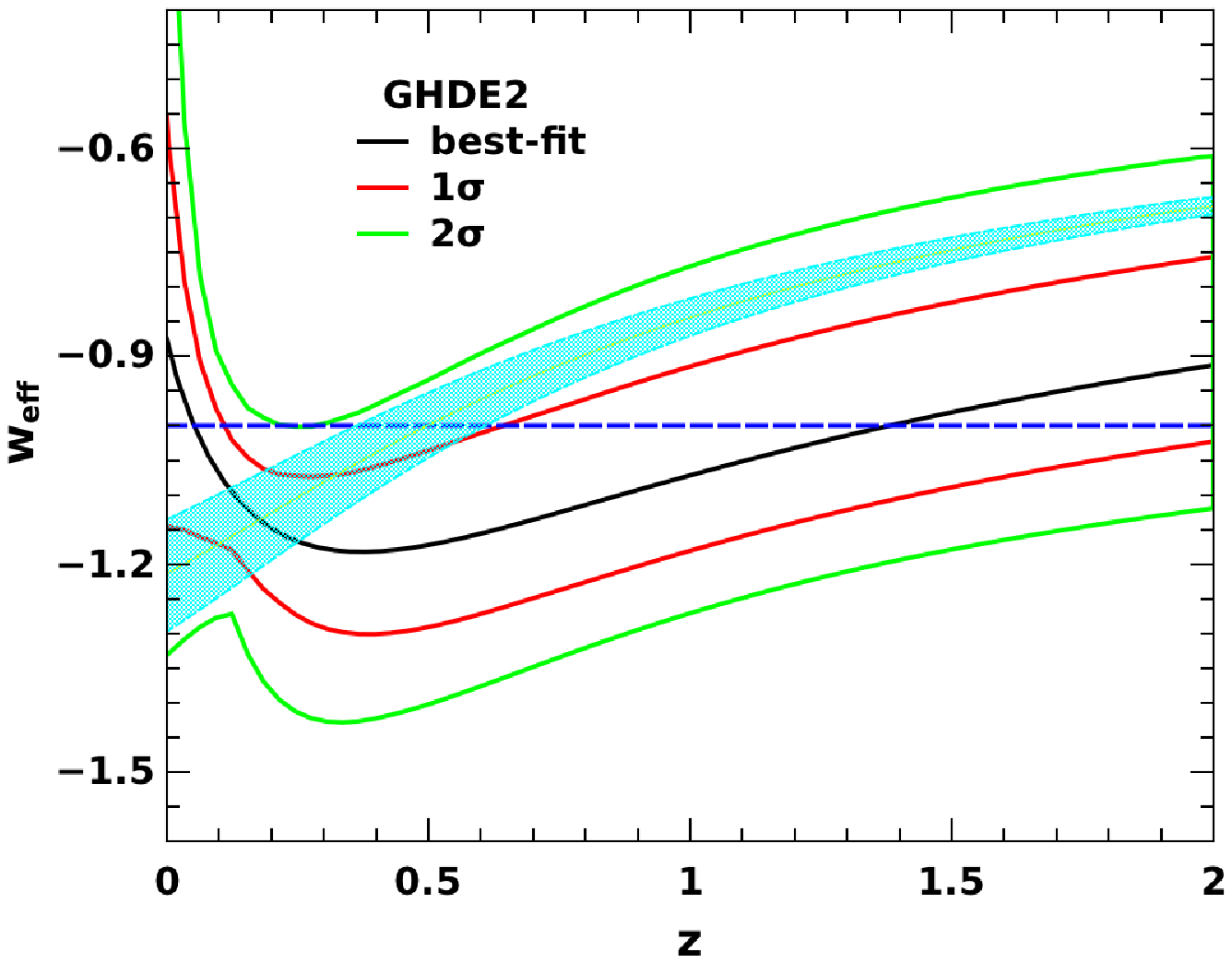}
\end{center}
\caption{ \label{figJBP}
$Left\ Panel$: Marginalized probability contours at 1$\sigma$ and 2$\sigma$ in the $c_0-c_1$ plane,
for the GHDE2 model.
The constraints Eq. (\ref{eq:GHDE2_condition}) are automatically satisfied.
$Right\ Panel$: The evolution of $w_{eff}$ along with $z$ in the GHDE2 model.
The $w=-1$ line and the 1$\sigma$ constraint of the original HDE model are also shown.
They are all consistent with the GHDE2 model in the 2$\sigma$ CL.}
\end{figure}

In the left panel of Fig.~\ref{figJBP}, we plot the 1$\sigma$ and 2$\sigma$ contours for the  GHDE2 model in the $c_0-c_1$ plane.
Again, the requirement of $\rho_{de}>0$ is automatically satisfied.
The evolution of $w_{eff}$ of this model is plotted in the right panel of Fig.~\ref{figJBP}.
Like the GHDE1 model, here we also found that both the $w_{eff}$ of the original HDE model and the cosmological constant roughly 
lie in the 2$\sigma$ error of the evolution of $w_{eff}$.

\subsection{GHDE3: the Wetterich type  }

The third parametrization considered is the ``Wetterich-type'' parametrization,
having the following form of $c(z)$,
\begin{equation}\label{eq:GHDE3c}
c(z)=\frac{c_0}{1+c_1\ln(1+z)}~.
\end{equation}
It has the property: $z\rightarrow 0,c\rightarrow c_0;~z\rightarrow +\infty,c\rightarrow 0.$
And it is also straightforward to have
\begin{equation}
\frac{(1+z)}{c}\frac{dc}{dz}=-\frac{c_1}{1+c_1\ln(1+z)}~.
\end{equation}
The condition $\rho_{de}>0$ reduces to,
\begin{equation}\label{eq:GHDE3_condition}
c_0>0~,\ \ c_1\geq0~.
\end{equation}

\begin{figure}
\begin{center}
\includegraphics[width=7.15cm]{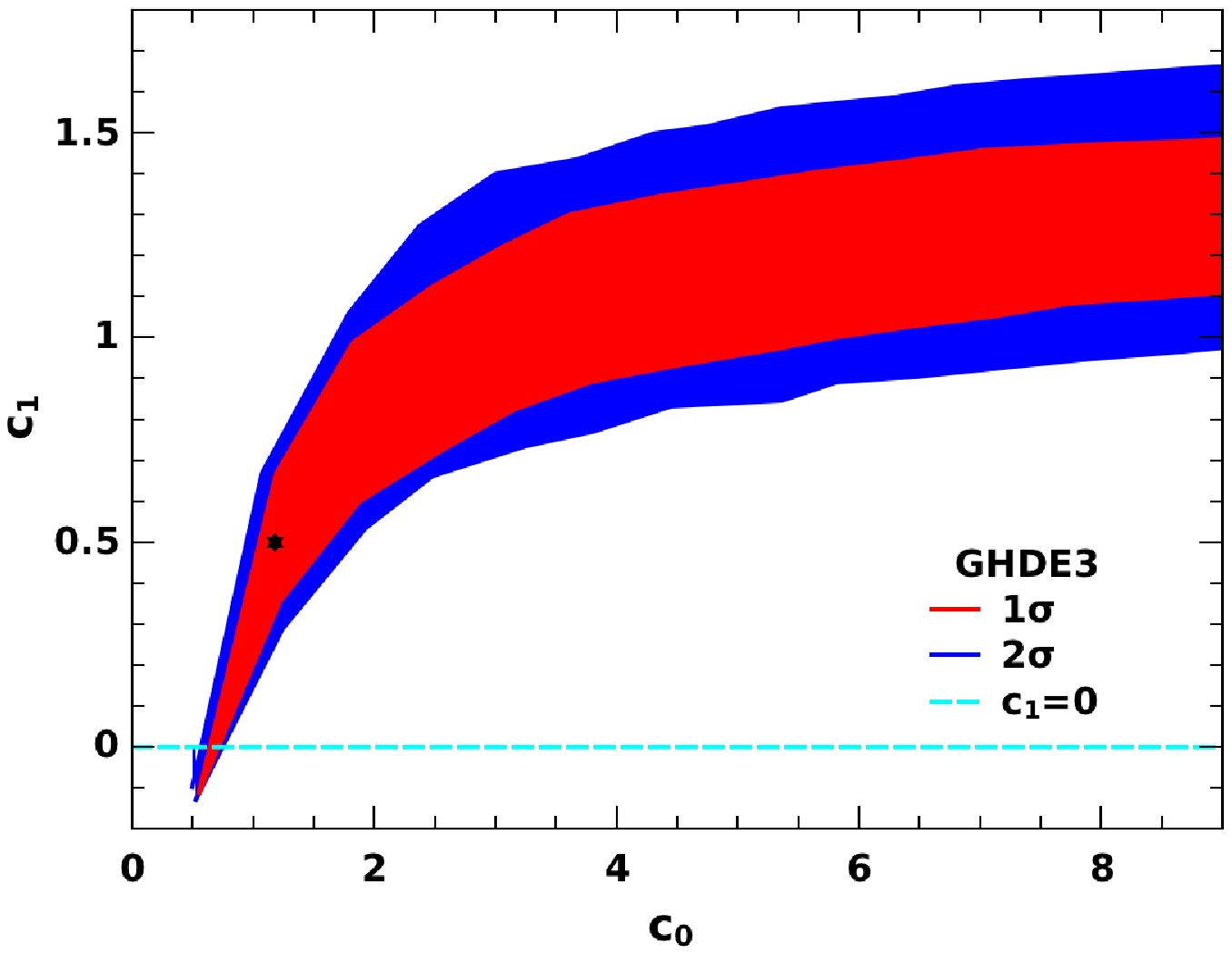}
\includegraphics[width=7.2cm]{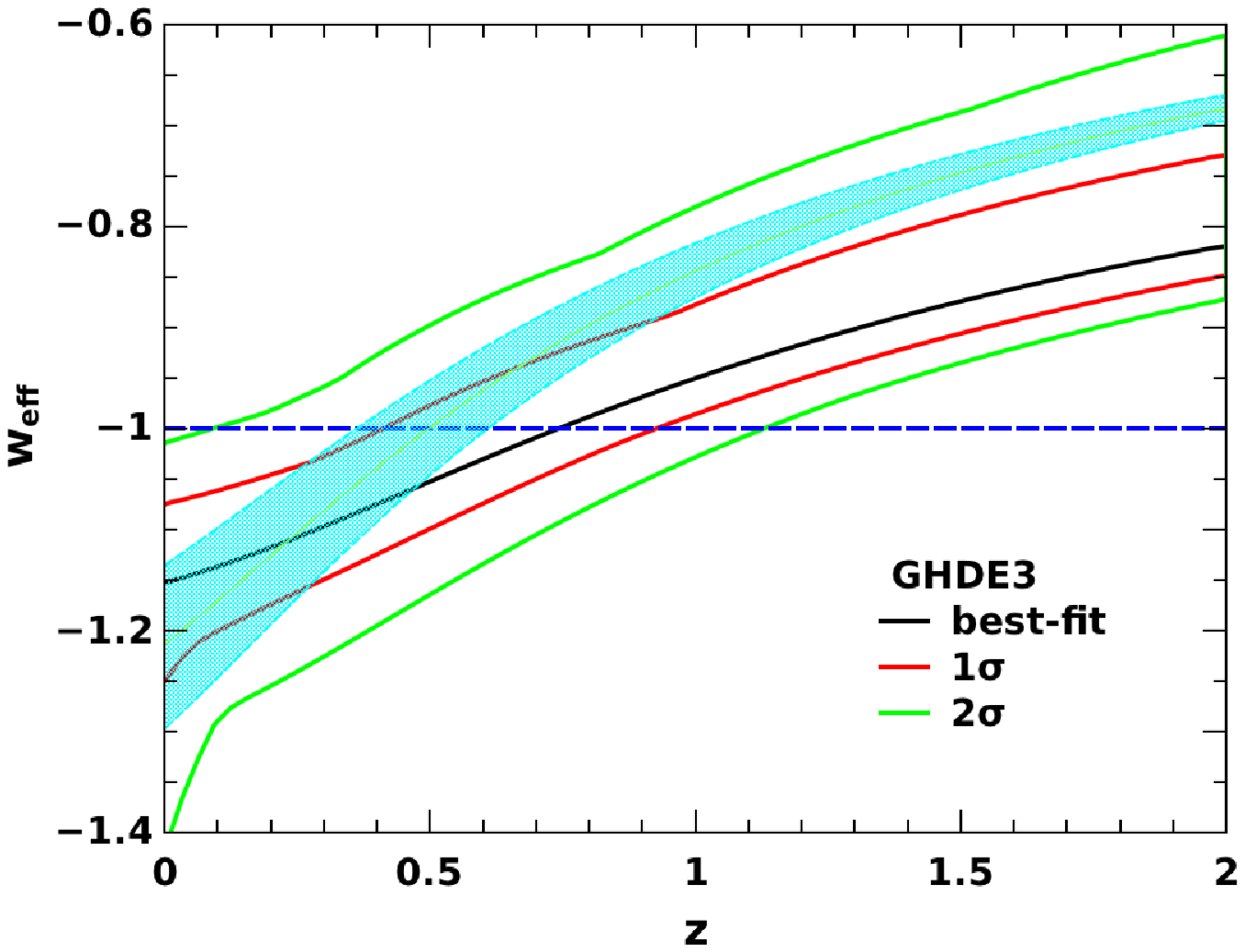}
\caption{\label{figHDEwet}
$Left\ Panel$: Marginalized probability contours at 1$\sigma$ and 2$\sigma$ CL in the $c_0-c_1$ plane,
for the GHDE3 model.
The constraints Eq. (\ref{eq:GHDE3_condition}) are slightly violated at small $c_0$.
$Right\ Panel$: The evolution of $w_{eff}$ along with $z$ in the GHDE3 model.
The cosmological constant is deviated at high redshift,
while the original HDE model is still consistent with this model in the 2$\sigma$ CL.}
\end{center}
\end{figure}

The contours of this model are plotted in the left panel of Fig.~\ref{figHDEwet}.
Here we find that the requirement of $\rho_{de}>0$ is slightly violated at the small value region of $c_0-c_1$ plane.
The evolution of  $w_{eff}$ of this model is plotted in the right panel of Fig.~\ref{figHDEwet}.
Unlike the GHDE1 and the GHDE2  model, here we found that  both at the very low~($z\lesssim0.1$) and high~($z\gtrsim1.2$) redshift regions, 
the behavior of GHDE deviates from the cosmological constant, 
while the original HDE model is still consistent with this GHDE model in the 2$\sigma$ CL.

\subsection{GHDE4: the Ma-Zhang type}\label{sec:log}

A common short-coming of the above three parametrizations is that they all diverge when the redshift $z$ approaches $-1$,
so we are unable to investigate the future behavior of GHDE in these models.
In this section, to overcome this, we consider the following form of $c(z)$, which is proposed in \cite{zhangxinlog}:
\begin{equation}\label{eq:GHDE3c}
c(z)=c_0+c_1\left(\frac{\ln(2+z)}{1+z}-\ln2\right)~.
\end{equation}
It follows directly that
\begin{eqnarray}
z\rightarrow 0:\ c\rightarrow c_0;\ \  z\rightarrow +\infty: \ c\rightarrow c_0-c_1\ln2;\ \ z\rightarrow -1:\ c\rightarrow c_0+(1-\ln2)c_1~,
\end{eqnarray}
\begin{equation}
\frac{(1+z)}{c}\frac{dc}{dz}=c_1\frac{\frac{1}{2+z}-\frac{\ln(2+z)}{1+z}}{c_0+c_1\left(\frac {\ln(2+z)}{1+z}-\ln2 \right)}~.
\end{equation}

We can find that the $c(z)$ of this parametrization is constrained in a finite value even in the future of the universe evolution.
Thus this model can have a well-defined property in the far future evolution of the universe, just as the original HDE model which possesses a constant c.

In this model, the requirement of $\rho_{de}>0$ is reduced to,
\begin{equation}\label{eq:GHDE4_condition}
c_0>0~,\ \ c_0-c_1\ln2>0~,\ \ c_0+(1-\ln2)c_1>0~.
\end{equation}

\begin{figure}
\begin{center}
\includegraphics[width=7.25cm]{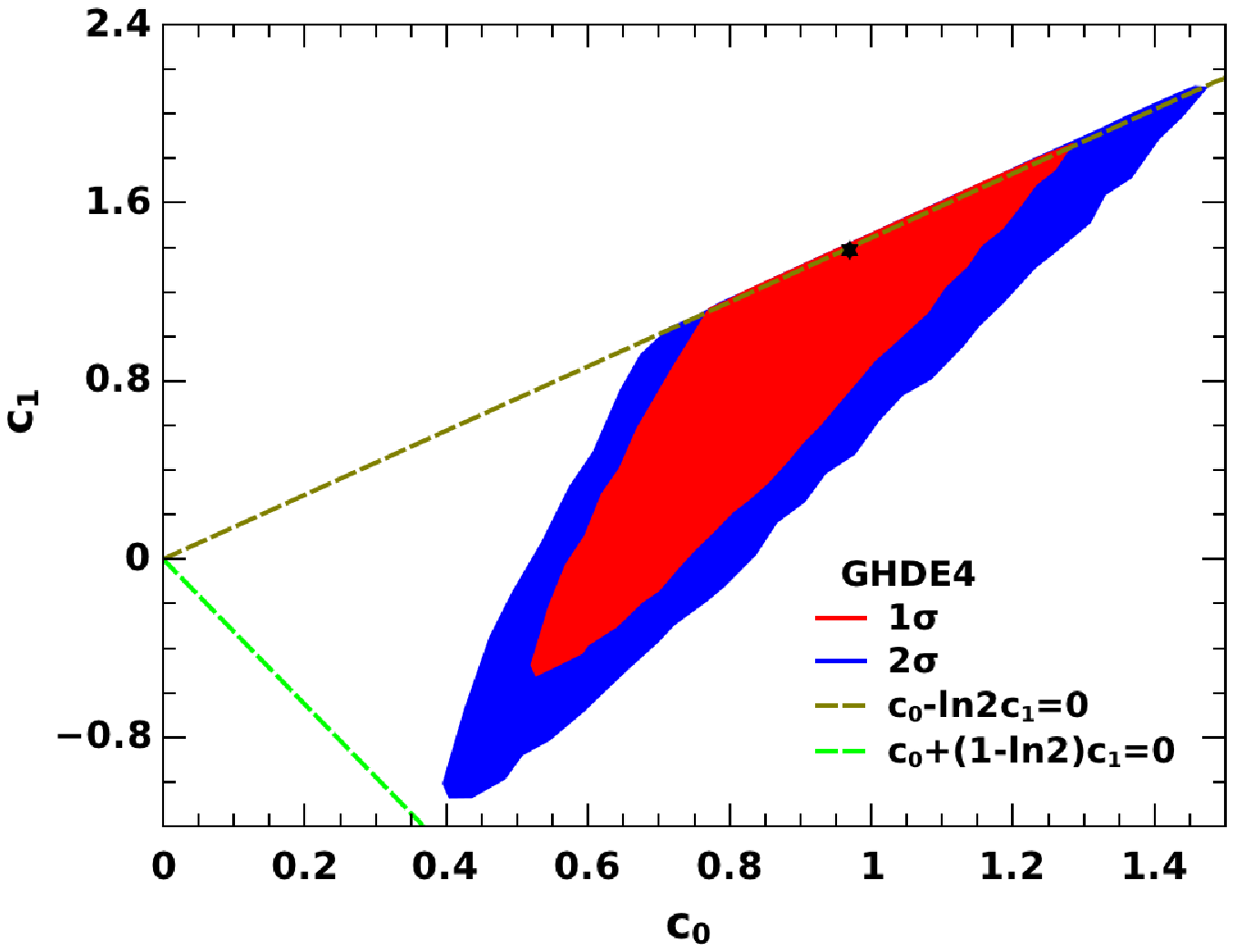}
\includegraphics[width=7.2cm]{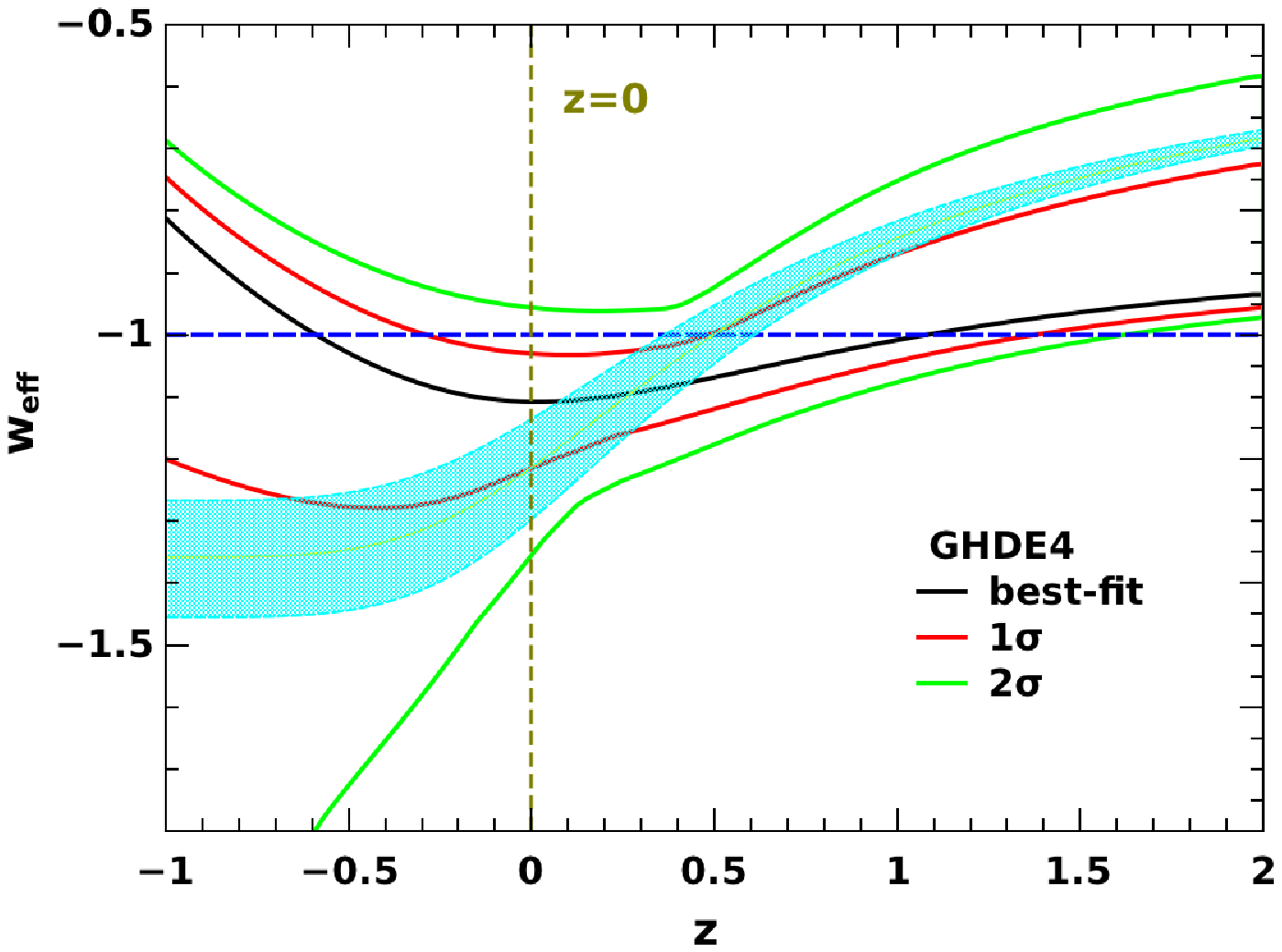}
\caption{\label{figlog}
$Left\ Panel$: Marginalized probability contours at 1$\sigma$ and 2$\sigma$ CL in the $c_0-c_1$ plane, for the GHDE4 model.
The physical conditions, Eq. (\ref{eq:GHDE4_condition}), are automatically satisfied.
$Right\ Panel$: The evolution of $w_{eff}$ along with $z$ in the GHDE4 model.
Different from the previous three GHDE models, the future evolution of $w_{eff}$ is also plotted.}
\end{center}
\end{figure}

In the left panel of Fig.~\ref{figlog}, we plot the 1$\sigma$ and 2$\sigma$ contours for the GHDE4 model in the $c_0-c_1$ plane.
The constraints of Eq. (\ref{eq:GHDE4_condition})
are automatically satisfied by the numerical simulation results.
We also plot the evolution of $w_{eff}(z)$ of this model in the right panel of Fig.~\ref{figlog}.
A most striking property we find in this model is that the future behavior of $w_{eff}(z)$ could be completely different:
while in the original model the behavior of DE is phantom-like in the future,
the GHDE in this model can behave like a quintessence, cosmological constant, or phantom.
All these possibilities are allowed according to the 2$\sigma$ constraint of the data (see Sec.~IV for detailed discussion).

\subsection{A  Summary of the GHDE Models Considered}

In the above subsections, we study four GHDE models with special $c(z)$s.
A brief summary of these models are shown in Table \ref{tabresults},
where the models, model parameters (together with their best-fit values and 1$\sigma$ uncertainties), and $\chi^2_{min}$s are given.
To make a comparison, the results of the original HDE model and the $\Lambda$CDM model (see Ref. \cite{SNLS3XDLi}) are also listed.
Here some nuisance parameters, such as $h$, $\alpha$ and $\beta$ mentioned in the appendix,
are not shown since they are not model parameters with significant meanings.

From the $\chi^2_{min}$, we can see that all these models can provide a nice fit to the observational data.
Especially, the GHDE2 model showed that the inclusion of the extra parameter can reduce the $\chi^2_{min}$ of the original HDE model by 2.16.

\begin{table} \caption{\label{tabresults} Fitting results for the GHDE models. }
\begin{center}
\label{table1}
\begin{tabular}{ccccc}
  \hline\hline
Model   &     $\Omega_{m0}$      &             $c_0$                 &          $c_1$        &     $\chi^2_{min}$    \\
  \hline
  $\Lambda$CDM ~&~~~ $0.265^{+0.014}_{-0.013}$     ~~&~~        $-$      ~~&~~     $-$       ~~&~~              424.911   ~~~            \\
  \hline
  HDE  ~&~~~ $0.261^{+0.014}_{-0.013}$ ~~&~~  $0.65^{+0.06}_{-0.06}   $   ~~&~~  $0 $ (fixed)       ~~&~~              424.855  ~~~            \\
  \hline
  GHDE1 ~&~~~ $0.265^{+0.012}_{-0.014}$ ~~&~~  $3.32^{+\ \ -}_{-2.45}$   ~~ &~~   $-2.94^{+2.66}_{-\ \ -}$~~& ~~ 423.369      ~~~  \\
  \hline
  GHDE2 ~&~~~ $0.266^{+0.015}_{-0.013}$ ~~&~~ $0.25^{+0.24}_{-0.12}$ ~~ & ~~     $ 0.65^{+0.16}_{-0.37}$ ~~&~~  422.696    ~~~  \\
  \hline
  GHDE3 ~&~~~ $0.262^{+0.013}_{-0.011}$ ~~&~~ $1.18^{+\ \ -}_{-0.44}$  ~~ & ~~  $ 0.50^{+0.66}_{-0.40}$ ~~&~~   423.735    ~~~  \\
  \hline
 GHDE4 ~&~~~ $0.265^{+0.012}_{-0.014}$ ~~&~~ $0.97^{+0.15}_{-0.28}$  ~~ & ~~  $1.38^{+0.24}_{-1.14}$    ~~& ~~  423.407   ~~~  \\
  \hline\hline
\end{tabular}
\end{center}
\end{table}

\begin{figure}
\makeatletter
          \def\@captype{figure}
          \makeatother
\begin{center}
\includegraphics[width=6.9cm]{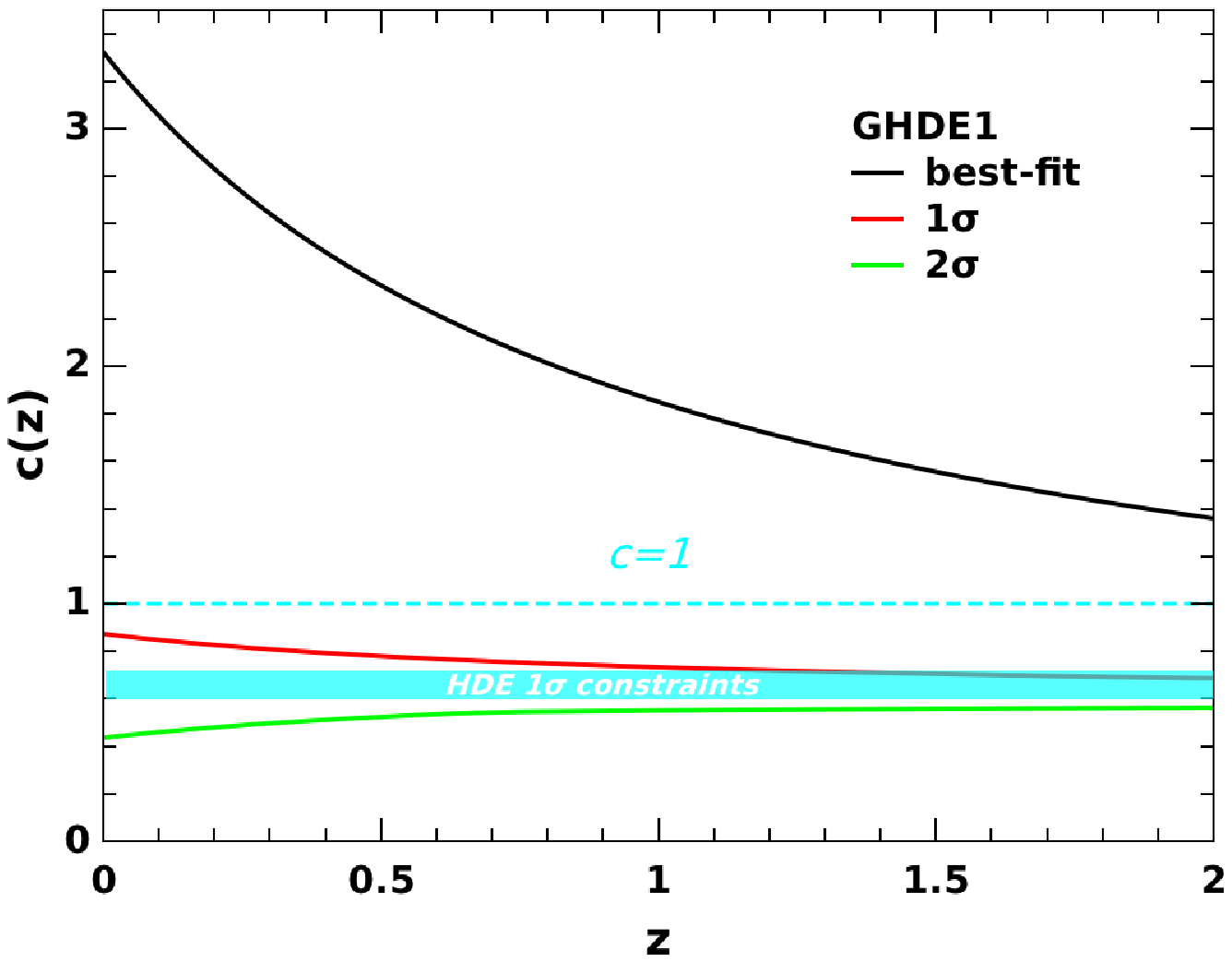}
\includegraphics[width=7.2cm]{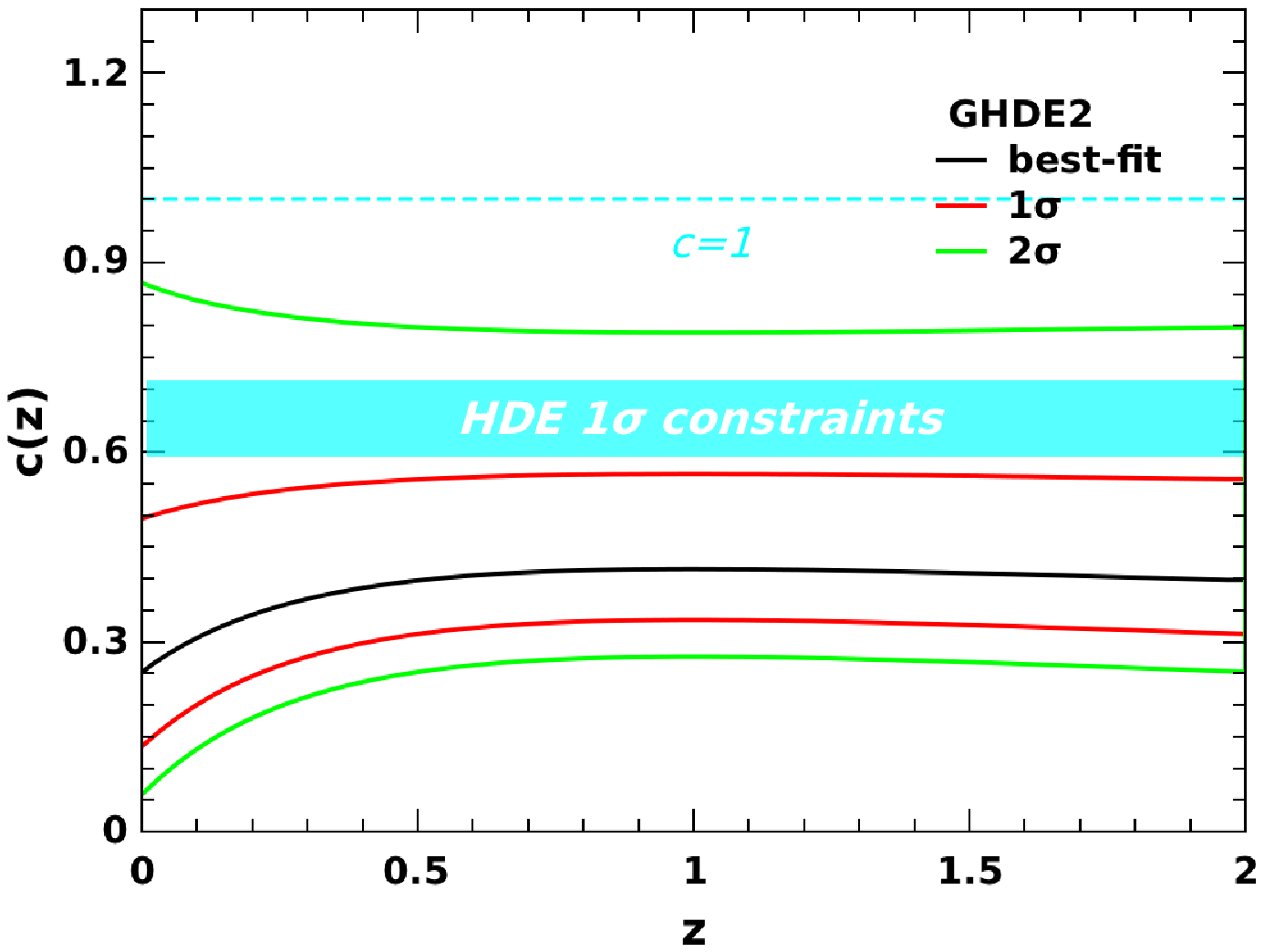}
\includegraphics[width=7.15cm]{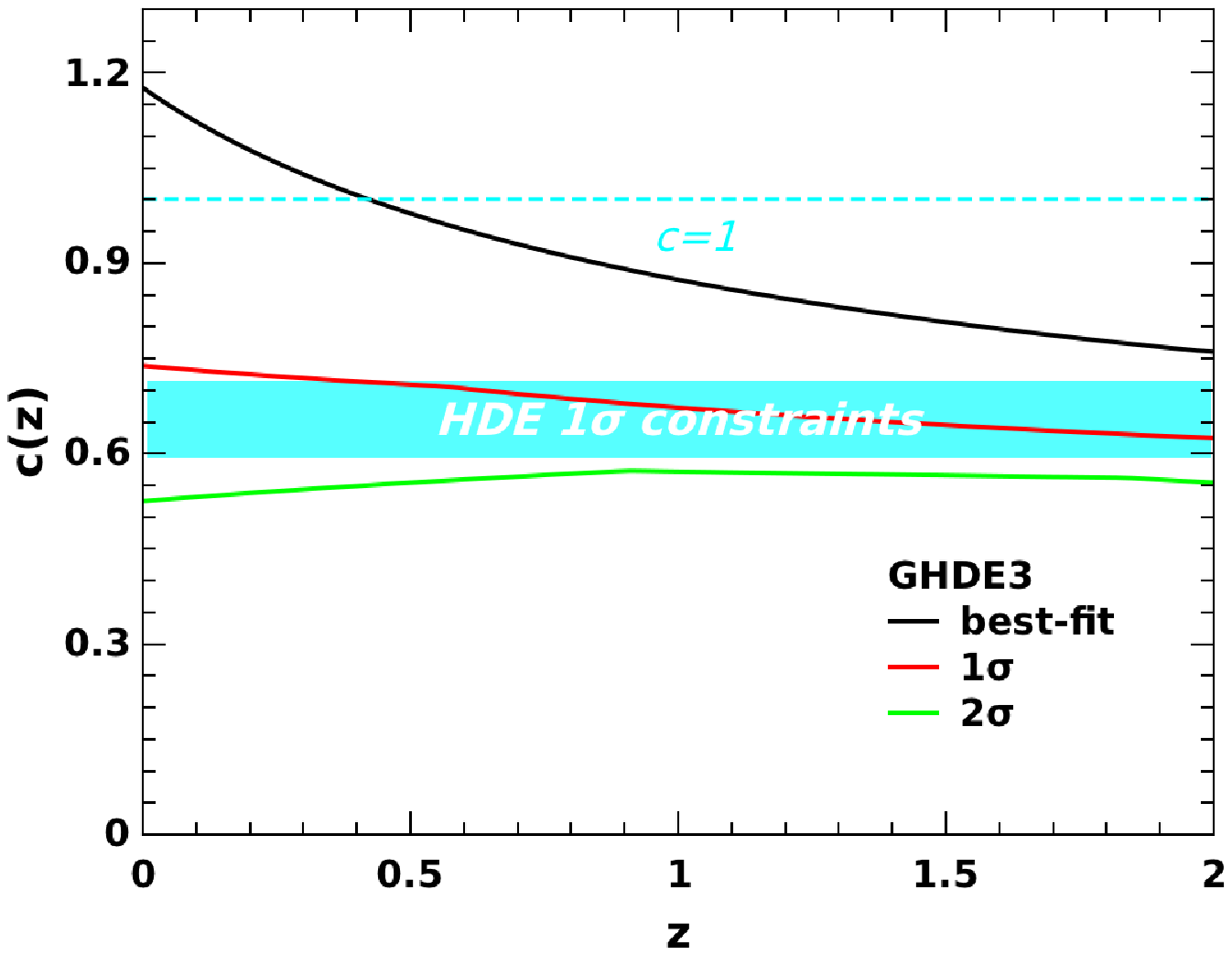}
\includegraphics[width=6.95cm]{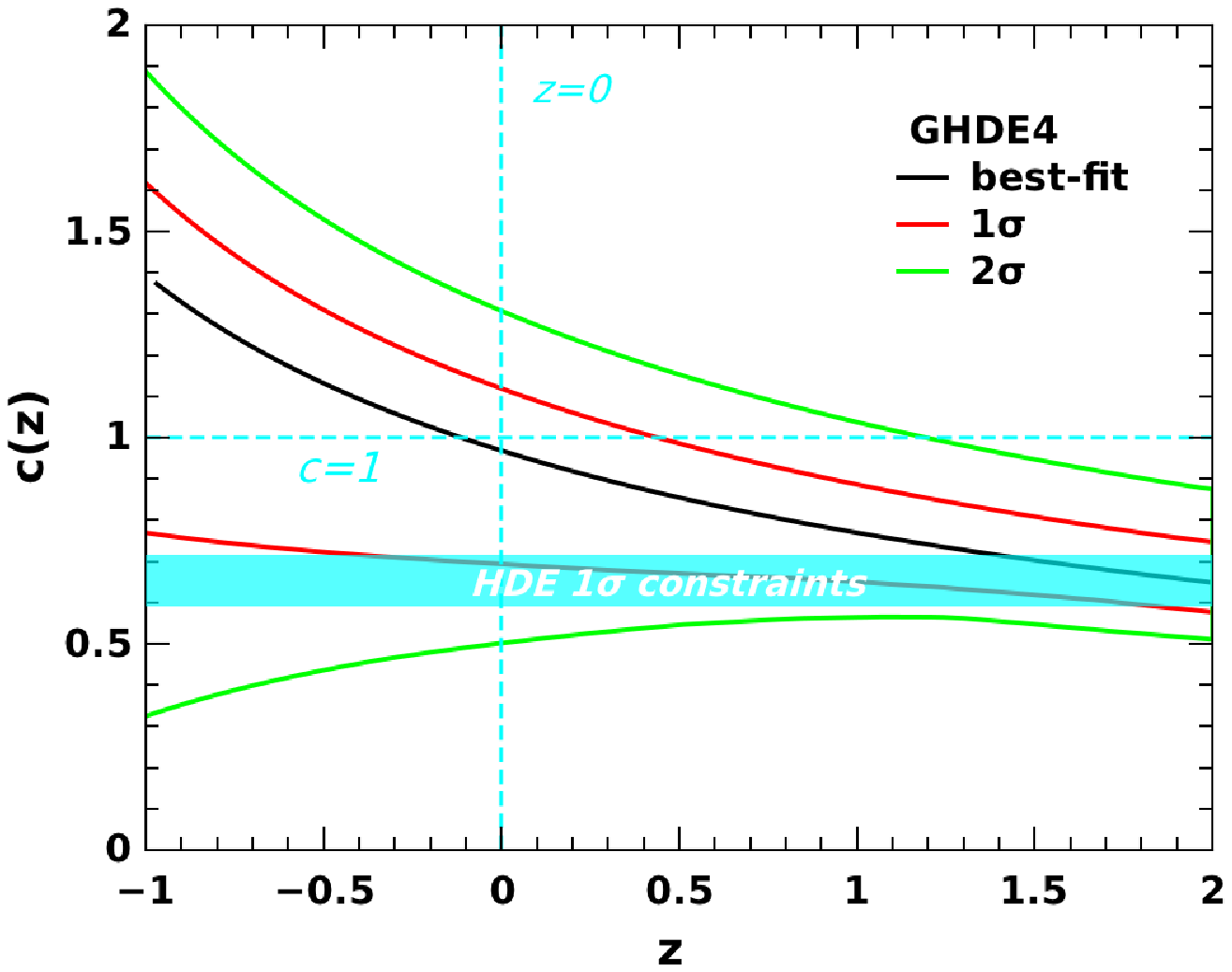}
\caption{\label{figGHDEcz}
 The evolution of $c$s along with $z$ for the four considered GHDE models. For having a loosely closed probability contours in the $c_0-c_1$ plane(see Fig.~\ref{figCPL} and Fig.~\ref{figJBP}), thus the upper bounds of $c(z)$ are missed in both the GHDE1 and GHDE3 models.}
\end{center}
\end{figure}

 By fitting the observational data, we also get the constraints for those four special $c(z)$ forms, which are plotted in Fig.~\ref{figGHDEcz}.
 It is found that both the best-fit and the constraint regions of $c(z)$ are different among the four GHDE models.
 As mentioned in Sec. \ref{sec:GHDE}, $\frac{dc(z)}{dz}$ plays an essential role in characterizing the evolution of DE in the GHDE models.
 Thus the difference can be interpreted by the different forms of $\frac{dc(z)}{dz}$s in these models.

We also find that, although all the  $c(z)$'s best-fits of these four GHDE models distinctly deviate from $c$'s 1$\sigma$ region of the original HDE model, 
the 1$\sigma$ constraint of the original HDE model lies well in the 2$\sigma$ region of each GHDE models considered in this work.
So we conclude that there are not enough information to determine the form of $c(z)$ in the GHDE scenario  based on the current cosmological observation data.

\section{The Fate of the Universe in the GHDE Scenario}

The fate of the universe is always a fascinating issue\cite{caldwell}. In this section, we discuss the fate of the universe in the GHDE scenario. First we study the fate in the original HDE model as a comparison.

\subsection{The fate in the original HDE model }

\begin{figure}
\begin{center}
\includegraphics[width=7.2cm]{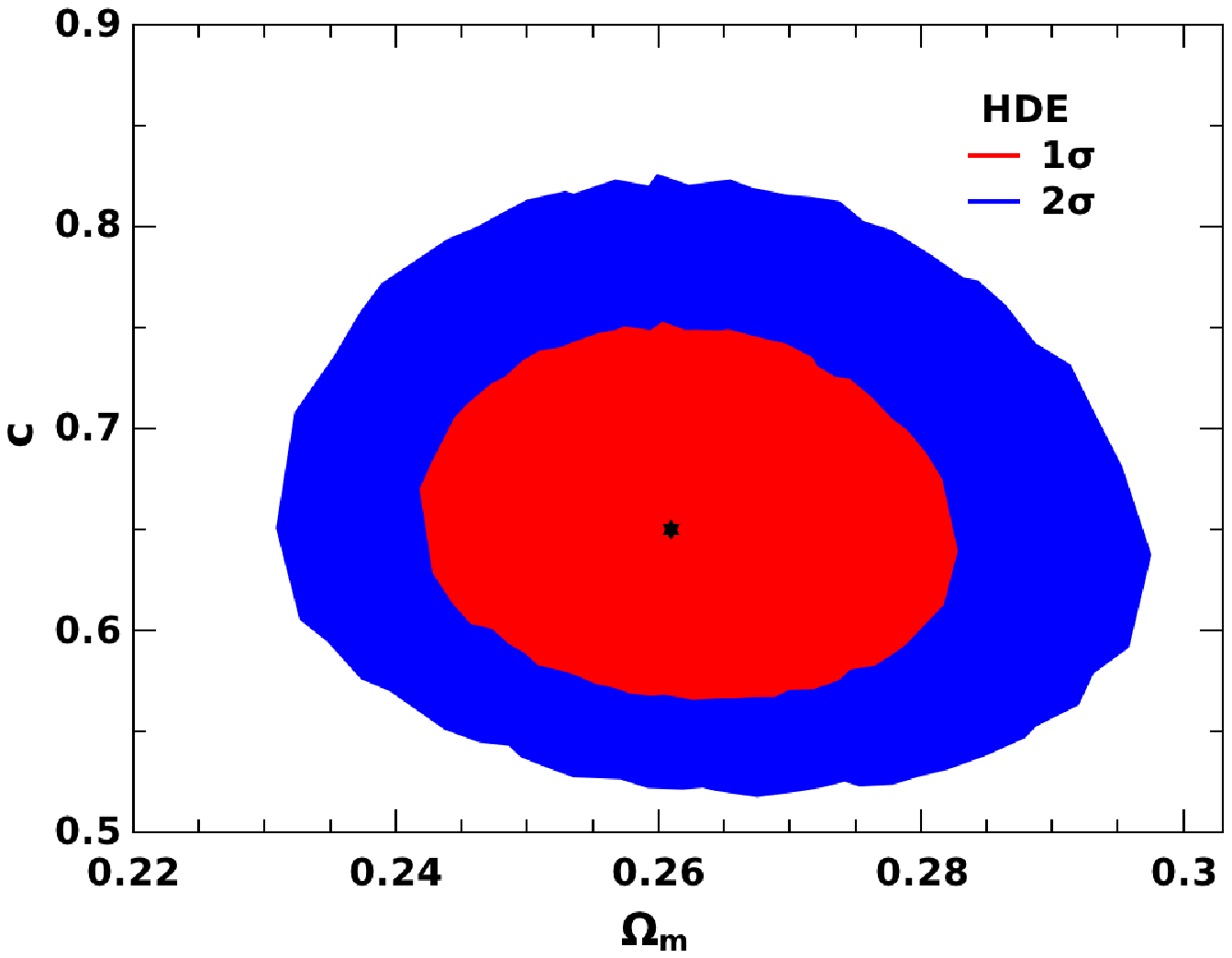}
\includegraphics[width=7.3cm]{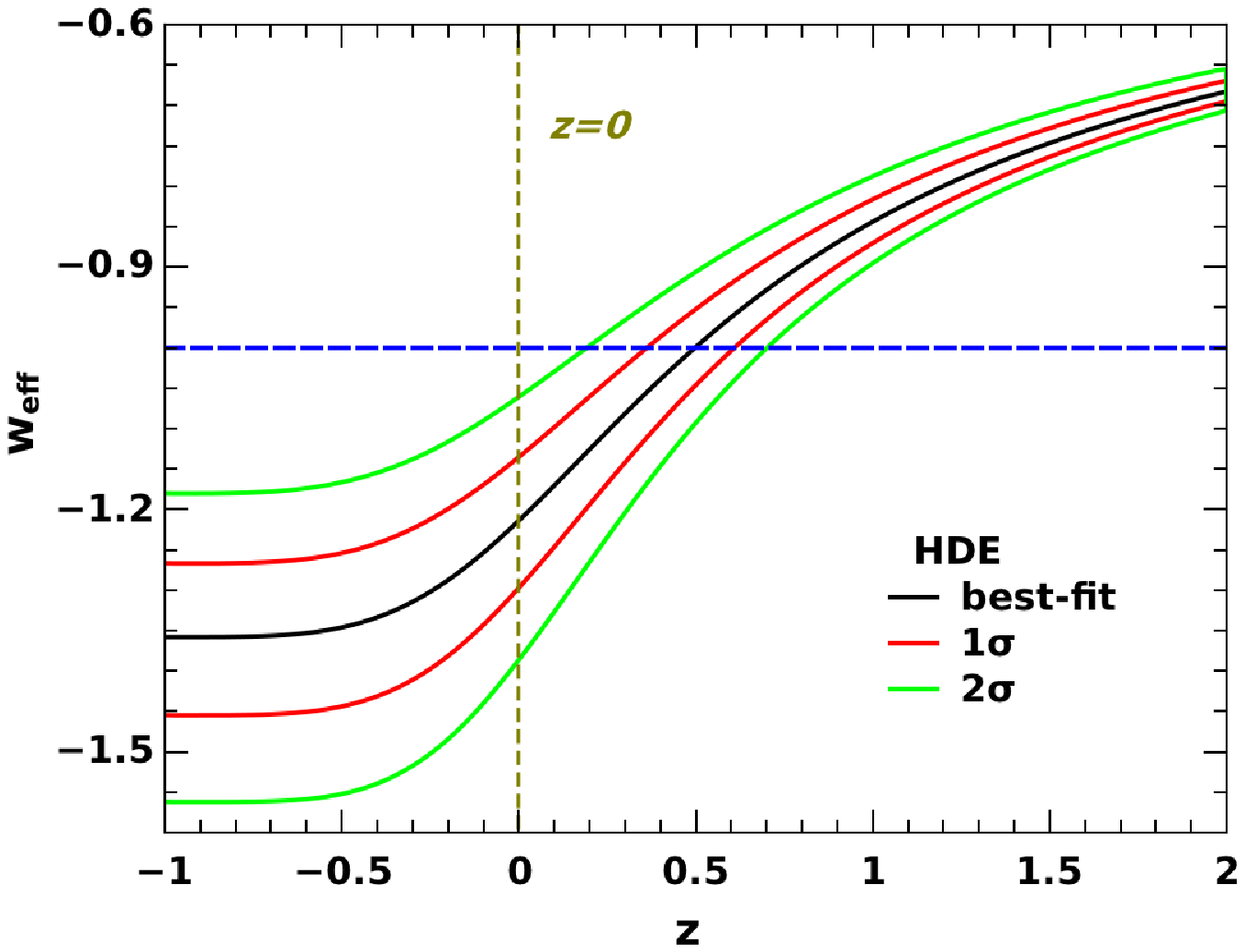}
\caption{\label{figHDE}
$Left\ Panel$: Marginalized probability contours at 1$\sigma$ and 2$\sigma$ CL in the $\Omega_m-c_0$ plane,
for the HDE model.
We obtain $c<1$ in the 2$\sigma$ CL, implying that the future behavior of HDE is phantom like.
$Right\ Panel$: The evolution of $w_{eff}$ along with $z$ for the HDE model.
The cosmological constant with $w=-1$ is also plotted.
In the far future when $z$ approaches -1, we have $w_{eff}<-1$ in the 2$\sigma$ CL.}
\end{center}
\end{figure}

As studied in Sec.~\ref{GHDE}, the behavior of HDE is essentially determined by the parameter $c$.  In the original HDE model, the parameter $c$ is proposed as a constant, thus the Eq. (\ref{eq:GHDEwde}) reduces to,
\begin{equation}\label{eq:HDEwde}
w_{de}=-{1\over 3}-{2\over 3c}\sqrt{\Omega_{de}}~.
\end{equation}
 From the discussion in Sec.~\ref{FRW}, we can find that in the far future, the DE will be the dominated composite of the universe, i.e.,  $z\rightarrow-1,~ \Omega_{de}\rightarrow1~$. From Eq. (\ref{eq:HDEwde}), we have,
\begin{equation}\label{eq:futurewde}
w_{de}\rightarrow -{1\over 3}-{2\over 3c},\ \ \ \ when\ \ z\rightarrow-1~.
\end{equation}
From Eq. (\ref{eq:futurewde}) the relation between $c$ and the fate of the universe can be seen clearly.
If $c=1$, in the far future HDE will behave like a cosmological constant;
if $c>1$, then always we have $w>-1$ and HDE behaves as a quintessence;
and if $c<1$, we will have $w<-1$ in the future,
leading to a phantom universe with big rip as its ultimate fate.

Based on the joint analysis from the SNLS3+BAO+CMB+$H_0$ data,
in the left panel of Fig.~\ref{figHDE}, the 1$\sigma$ and 2$\sigma$ contours for the HDE model in the $\Omega_m-c_0$ plane are plotted.
In the 2$\sigma$ CL, we obtain
\begin{equation}
\Omega_m =0.265^{+0.029}_{-0.025} ,\ \ \ c =0.65^{+0.14}_{-0.11} ,
\end{equation}
showing that $c<1$ is significantly favored by the data,
which implies that, for the orginal HDE model, there will be $w_{eff}<-1$ in the far future
(see the right panel of Fig.~\ref{figHDE}).
Thus, in the orginal HDE model the behavior of HDE is phantom like at last,
and the universe will probably end up with a big rip\cite{caldwell}.

\subsection{The fate in the GHDE models}

In the GHDE scenario, the parameter $c$ is not a constant but a function of the redshit $z$. This makes the variety of the HDE behavior increases, and thus let the GHDE model present many more different possibilities for the fate of the universe.

We can focus on the GHDE4 model for a typical demonstration.
From Fig.~\ref{figlog},
 it is interesting to see that various possibilities emerge even in 1$\sigma$ region:
the behavior of GHDE in the future evolution can be either quintessence-like or phantom-like, which means the big rip can either happen or not.

Inspired by this fact, one may wonder what forms of $c(z)$ can make the HDE processing a purely quintessence-like, phantom-like, or even cosmological-constant-like behavior.
This issue is to some extent equivalent with another issue,
that is,  how to reconstruct a given form of DE component in the GHDE scenario.

For a most general form of DE component, we can describe it by the DE density function $f(z)$,
\begin{equation}
\rho_{de}(z)=\rho_{de0}f(z)~.
\end{equation}
 Combining with Eqs.~(\ref{eq:rh}) and (\ref{eq:rhoGHDE}), we have
\begin{equation}
c(z)= H_0R_h\sqrt{\Omega_{de0}f(z)} =\frac{\sqrt{(1-\Omega_{m0}-\Omega_{r0})f(z)}}{1+z}\int^z_{-1}{dz^\prime\over E(z^\prime)}~.
\end{equation}
So indeed we can utilizing the GHDE model to reconstruct a DE component with arbitrary from,
once given its energy density function $f(z)$.

As an example, let us consider the case that the DE component with a constant $w$, i.e. $f(z)=(1+z)^{3(1+w)}$. Neglecting the small $\Omega_r$, thus
\begin{equation}\label{eq:constantw}
c(z)=\sqrt{(1-\Omega_{m0})(1+z)^{1+3w}}\int^z_{-1}{dz^\prime\over\sqrt{\Omega_{m0}(1+z^\prime)^3+(1-\Omega_{m0})(1+z^\prime)^{3(1+w)}}}.
\end{equation}

\begin{figure}
\begin{center}
\includegraphics[width=7cm]{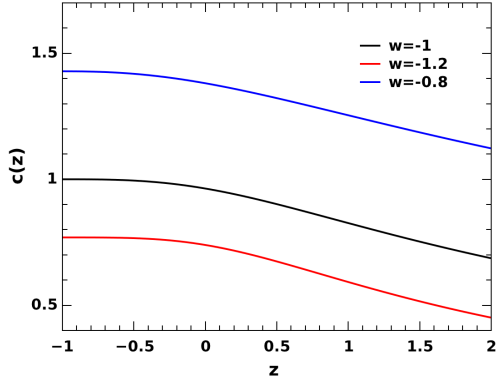}
\end{center}
\caption{\label{figconstantwz} The evolution of $c$  along with $z$ that corresponding to different value of constant $w$.
especially $w=-1$ is the cosmological constant case.}
\end{figure}

A demonstration is shown in Fig.~\ref{figconstantwz}.
 We take the value $\Omega_{m0}=0.25$ and $w=-0.8,-1.2,-1$, which correspond to a purely quintessence-like, phantom-like, and cosmological constant behavior, respectively.
By numerically integrating the Eq. (\ref{eq:constantw}),
we get the evolution of $c(z)$ for these $w$s.
Especially, the case $w=-1$ gives the well-known $\Lambda$CDM model.
Thus, by choosing different forms of $c(z)$s,
we can reconstruct a DE component with various behaviors.

As already discussed in the above section, that we can not confirm the form of $c(z)$  based on the current cosmological observation data.
So, in the GHDE scenario, the fate of the universe remains a mystery.

\section{Concluding Remarks}\label{Results}

In this work, we investigate the generalized holographic dark energy models by considering a redshift-dependent $c(z)$.
In all, we consider four parametrizations of $c(z)$,
including the CPL type, JBP type, Wetterich type and Ma-Zhang type parametrizations.
By performing the joint data analysis, it is shown that all these models can give a nice fit to the data.
Among them, the JBP-type GHDE model can even lead to a reduction in the $\chi^2_{min}$ of 2.16.

From the fitting results we also find that the generalization of $c(z)$ increase the variety of the HDE model.
However, the evolutions of $w_{eff}(z)$ in these models are all consistent with that of the original HDE model in 2$\sigma$ CL.  
That means, we do not see clear evidence for the evolution of $c(z)$ based on the current observational data.

The fate of the universe is always a fascinating issue.
Unlike the original HDE model predicting a probably big-rip in the far future,
we find that the GHDE model, with the form of $c(z)$ undetermined by the current data,
lay this issue back to a mystery.
Thus for the GHDE model, big-rip or not, it continues to be a question.

At the end, we hope that the method of considering the parameter $c$ of the HDE as a redshift-dependent function
may provide us some clues to understand the DE problem.
In all, more studies about this scenario are needed in the future.

\section*{Acknowledgements}
We would like to thank Xin Zhang for helpful discussions.
This work was supported by the NSFC under Grant Nos.
10535060, 10975172 and 10821504,
and by the 973 program (Grant No. 2007CB815401) of the Ministry of Science and Technology of China.

\

\

\

\

\section{Appendix: Observational Data and Methodology}\label{Data}

In this work, we adopt the $\chi^2$ statistic to estimate the model parameters.
For a physical quantity $\xi$ with experimentally measured value $\xi_{obs}$,
standard deviation $\sigma_{\xi}$ and theoretically predicted value $\xi_{th}$,
the $\chi^2$ takes the form
\begin{equation}
\chi^2_{\xi}={(\xi_{obs}-\xi_{th})^2\over \sigma^2_{\xi}}~.
\end{equation}
The total $\chi^2$ is the sum of all $\chi^2_{\xi}$s, i.e.
\begin{equation}
\chi^2=\sum_{\xi}\chi^2_{\xi}~.
\end{equation}
One can determine the best-fit model parameters by minimizing the total $\chi^2$.
Moreover, by calculating $\Delta \chi^2 \equiv \chi^2-\chi^2_{\rm min}$,
one can determine the 1$\sigma$ and the 2$\sigma$ confidence level (CL) ranges of a specific model.
Statistically, for models with different $n_p$ (denoting the number of free model parameters),
the 1$\sigma$ and 2$\sigma$ CL correspond to different $\Delta \chi^2$.
In Table \ref{NpDchisq}, we list the relationship between $n_p$ and $\Delta \chi^2$ from $n_p=1$ to $n_p=9$.

In this work,
we determine the best-fit parameters and the 1$\sigma$ and 2$\sigma$ CL ranges by using the Monte Carlo Markov chain (MCMC) technique.
We modify the publicly available CosmoMC package \cite{COSMOMC} and generate $O(10^6)$ samples for each set of results presented in this paper.
We also verify the reliability and accuracy of the code by using the Mathematica program \cite{Wolfram}.

\begin{table}
\caption{Relationship between number of free model parameters $n_p$ and $\Delta \chi^2$.}
\begin{center}
\label{NpDchisq}
\begin{tabular}{|c|c|c|}
  \hline
  ~~~$n_p$~~~ & ~~~$\Delta \chi^2(1\sigma)$~~~ & ~~~$\Delta \chi^2(2\sigma)$~~~ \\
  \hline
  ~~~  1  ~~~ &          ~~~  $1$  ~~~        &          ~~~  $4$  ~~~          \\
  \hline
  ~~~  2  ~~~ &          ~~~$2.30$ ~~~        &          ~~~$6.18$ ~~~          \\
  \hline
  ~~~  3  ~~~ &          ~~~$3.53$ ~~~        &          ~~~$8.02$ ~~~          \\
  \hline
  ~~~  4  ~~~ &          ~~~$4.72$ ~~~        &          ~~~$9.72$ ~~~          \\
  \hline
  ~~~  5  ~~~ &          ~~~$5.89$ ~~~        &          ~~~$11.31$ ~~~         \\
  \hline
  ~~~  6  ~~~ &          ~~~$7.04$ ~~~        &          ~~~$12.85$ ~~~         \\
  \hline
  ~~~  7  ~~~ &          ~~~$8.18$ ~~~        &          ~~~$14.34$ ~~~         \\
  \hline
  ~~~  8  ~~~ &          ~~~$9.30$ ~~~        &          ~~~$15.79$ ~~~         \\
  \hline
  ~~~  9  ~~~ &          ~~~$10.42$ ~~~        &          ~~~$17.21$ ~~~         \\
  \hline
\end{tabular}
\end{center}
\end{table}

For data, we use the SNLS3 SNIa sample \cite{SNLS3},
the CMB anisotropy data from the WMAP7 observations \cite{WMAP7},
the BAO results from the SDSS DR7 \cite{SDSSDR7},
and the Hubble constant measurement from the WFC3 on the HST \cite{HSTWFC3}.
In the following, we briefly describe how these data are included into the $\chi^2$ analysis.

\subsection{The SNIa data}

Here we use the SNLS3 SNIa dataset released in \cite{SNLS3}.
This combined sample consists of 472 SN at $0.01 < z < 1.4$,
including 242 SN over $0.08 < z < 1.06$ from SNLS 3-yr observations \cite{SNLS3},
123 SN at low redshifts \cite{lowzSNIa}\cite{Constitution}, 93 SN at intermediate redshifts from the SDSS-II SN search \cite{SDSS-II},
and 14 SN at $z > 0.8$ from HST \cite{HSTSNIa}.
The systematic uncertainties of the SNIa data were nicely handled \cite{SNLS3}.
The total data of the SNLS3 sample can be downloaded from \cite{SNLS3Code}.

The $\chi^2$ of the SNIa data is
\begin{equation}
\chi^2_{\rm SN}=\Delta \overrightarrow{\bf m}^T \cdot {\bf C}^{-1} \cdot \Delta \overrightarrow{\bf m},
\end{equation}
where {\bf C} is a $472 \times 472$ covariance matrix capturing the statistic and systematic uncertainties of the SNIa sample,
and $\Delta {\overrightarrow {\bf m}} = {\overrightarrow {\bf m}}_B - {\overrightarrow {\bf m}}_{\rm mod}$ is a vector of model residuals of the SNIa sample.
Here $m_B$ is the rest-frame peak $B$ band magnitude of the SNIa,
and $m_{\rm mod}$ is the predicted magnitude of the SNIa given by the cosmological model
and two other quantities (stretch and color) describing the light-curve of the particular SNIa.
The model magnitude $m_{\rm mod}$ is given by
\begin{equation}\label{SNchisq}
m_{\rm mod} = 5\log_{10} \mathcal{D}_L(z_{\rm hel},z_{\rm cmb})-\alpha(s-1)+\beta \mathcal{C} + \mathcal{M}~.
\end{equation}
Here $\mathcal{D}_L$ is the Hubble-constant free luminosity distance, which takes the form
\begin{equation}
\mathcal{D}_L(z_{\rm hel}, z_{\rm zcmb}) = (1+z_{\rm hel})\int^{z_{\rm cmb}}_0 {{d z^\prime}\over{E(z^\prime)}}~,
\end{equation}
where $z_{\rm cmb}$ and $z_{\rm hel}$ are the CMB frame and heliocentric redshifts of the SN,
$s$ is the stretch measure for the SN,
and $\mathcal{C}$ is the color measure for the SN.
$\alpha$ and $\beta$ are nuisance parameters which characterize the
stretch-luminosity and color-luminosity relationships, respectively.
Following \cite{SNLS3}, we treat $\alpha$ and $\beta$ as free parameters and let them run freely.

The quantity $\mathcal{M}$ in Eq. (\ref{SNchisq}) is a nuisance parameter
representing some combination of the absolute magnitude of a fiducial SNIa and the Hubble constant.
In this work, we marginalize $\mathcal{M}$ following the complicated formula in the Appendix C of \cite{SNLS3}.
This procedure includes the host-galaxy information \cite{SullivanHostGalaxy} in the cosmological fits
by splitting the samples into two parts and allowing the absolute magnitude to be different between these two parts.

The total covariance matrix {\bf C} in Eq. (\ref{SNchisq}) captures both the statistical and systematic uncertainties of the SNIa data.
One can decompose it as \cite{SNLS3},
\begin{equation}
{\bf C} = {\bf D}_{\rm stat} + {\bf C}_{\rm stat} + {\bf C}_{\rm sys}~,
\end{equation}
where ${\bf D}_{stat}$ is the purely diagonal part of the statistical uncertainties,
${\bf C}_{\rm stat}$ is the off-diagonal part of the statistical uncertainties,
and ${\bf C}_{\rm sys}$ is the part capturing the systematic uncertainties.
It should be mentioned that, for different $\alpha$ and $\beta$, these covariance matrices are also different.
Therefore, in practice one has to reconstruct the covariance matrix $\bf C$ for the corresponding values of $\alpha$ and $\beta$,
and calculate its inversion.
For simplicity, we do not describe these covariance matrices one by one.
One can refer to the original paper \cite{SNLS3} and the public code \cite{SNLS3Code}
for more details about the explicit forms of the covariance matrices and the details of the calculation of $\chi^2_{\rm SN}$.

\subsection{The CMB data}

Here we use the ``WMAP distance priors'' given by the 7-yr WMAP observations \cite{WMAP7}.
The distance priors include the ``acoustic scale'' $l_A$, the ``shift parameter'' $R$, and the redshift of the decoupling epoch of photons $z_*$.
The acoustic scale $l_A$, which represents the CMB multipole corresponding to the location of the acoustic peak,
is defined as \cite{WMAP7}
\begin{equation}
\label{ladefeq} l_A\equiv (1+z_*){\pi D_A(z_*)\over r_s(z_*)}~.
\end{equation}
Here $D_A(z)$ is the proper angular diameter distance, given by
\begin{equation}
D_A(z)=\frac{1}{1+z}\int^z_0\frac{dz^\prime}{H(z^\prime)}~,
\label{eq:da}
\end{equation}
and $r_s(z)$ is the comoving sound horizon size, given by
\begin{equation}
r_s(z)=\frac{1} {\sqrt{3}}  \int_0^{1/(1+z)}  \frac{ da } { a^2H(a)
\sqrt{1+(3\Omega_{b}/4\Omega_{\gamma})a} }~,
\label{eq:rs}
\end{equation}
where $\Omega_{b}$ and $\Omega_{\gamma}$ are the present baryon and photon density parameters, respectively.
In this paper, we adopt the best-fit values, $\Omega_{b}=0.02253 h^{-2}$ and $\Omega_{\gamma}=2.469\times10^{-5}h^{-2}$ (for $T_{cmb}=2.725$ K),
given by the 7-yr WMAP observations \cite{WMAP7}.
The fitting function of $z_*$ was proposed by Hu and Sugiyama \cite{Hu:1995en}:
\begin{equation}
\label{zstareq} z_*=1048[1+0.00124(\Omega_b
h^2)^{-0.738}][1+g_1(\Omega_m h^2)^{g_2}]~,
\end{equation}
where
\begin{equation}
g_1=\frac{0.0783(\Omega_b h^2)^{-0.238}}{1+39.5(\Omega_b h^2)^{0.763}}~,
\quad g_2=\frac{0.560}{1+21.1(\Omega_b h^2)^{1.81}}~.
\end{equation}
In addition, the shift parameter $R$ is defined as \cite{Bond97}
\begin{equation}
\label{shift} R(z_*)\equiv \sqrt{\Omega_m H_0^2}(1+z_*)D_A(z_*)~.
\end{equation}
This parameter has been widely used to constrain various cosmological models \cite{Add3}.

As shown in \cite{WMAP7}, the $\chi^2$ of the CMB data is
\begin{equation}
\chi_{CMB}^2=(x^{obs}_i-x^{th}_i)(C_{CMB}^{-1})_{ij}(x^{obs}_j-x^{th}_j),\label{chicmb}~,
\end{equation}
where $x_i=(l_A, R, z_*)$ is a vector,
and $(C_{CMB}^{-1})_{ij}$ is the inverse covariance matrix.
The 7-yr WMAP observations \cite{WMAP7} had given the maximum likelihood values:
$l_A(z_*)=302.09$, $R(z_*)=1.725$, and $z_*=1091.3$.
The inverse covariance matrix was also given in \cite{WMAP7}
\begin{equation}
(C_{CMB}^{-1})=\left(
  \begin{array}{ccc}
    2.305 & 29.698 & -1.333 \\
    29.698 & 6825.27 & -113.180 \\
    -1.333 & -113.180  &  3.414 \\
  \end{array}
\right).
\end{equation}

\subsection{The BAO data}

Here we use the distance measures from the SDSS DR7 \cite{SDSSDR7}.
One effective distance measure is the $D_V(z)$, which can be obtained from the spherical average \cite{Eisenstein}
\begin{equation}
 D_V(z) \equiv \left[(1+z)^2D_A^2(z)\frac{z}{H(z)}\right]^{1/3},
\end{equation}
where $D_A(z)$ is the proper angular diameter distance.
In this work we use two quantities $d_{0.2}\equiv r_s(z_d)/D_V(0.2)$ and $d_{0.35}\equiv r_s(z_d)/D_V(0.35)$.
The expression of $r_s$ is given in Eq.(\ref{eq:rs}),
and $z_d$ denotes the redshift of the drag epoch, whose fitting formula is proposed by Eisenstein and Hu \cite{BAODefzd}
\begin{equation}
\label{Defzd} z_d={1291(\Omega_mh^2)^{0.251}\over 1+0.659(\Omega_mh^2)^{0.828}}\left[1+b_1(\Omega_bh^2)^{b2}\right]~,
\end{equation}
where
\begin{eqnarray}\label{Defb1b2}
b_1 &=& 0.313(\Omega_mh^2)^{-0.419}\left[1+0.607(\Omega_mh^2)^{0.674}\right], \\
b_2 &=& 0.238(\Omega_mh^2)^{0.223}.
\end{eqnarray}
Following \cite{SDSSDR7}, we write the $\chi^2$ for the BAO data as,
\begin{equation}
\chi^2_{BAO}=\Delta{p_i}(C_{BAO}^{-1})_{ij}\Delta p_j,
\end{equation}
where
\begin{equation}
\Delta p_i = p^{\rm data}_i - p_i,
\ \  p^{\rm data}_1 = d^{\rm data}_{0.2 } = 0.1905,
\ \  p^{\rm data}_2 = d^{\rm data}_{0.35} = 0.1097,
\end{equation}
and the inverse covariance matrix takes the form
\begin{equation}
(C_{BAO}^{-1})=\left(
  \begin{array}{cc}
    30124  & -17227 \\
    -17227 & 86977 \\
  \end{array}
\right).
\end{equation}

\subsection{The Hubble constant data}

The precise measurements of $H_0$ will be helpful to break the degeneracy between it and the DE parameters \cite{H0Freedman}.
When combined with the CMB measurement, it can lead to precise mesure of the DE EOS $w$ \cite{H0WHu}.
Recently, using the WFC3 on the HST, Riess {\it et al.} obtained an accurate determination of the Hubble constant \cite{HSTWFC3}
\begin{equation}
H_0=73.8\pm 2.4 {\rm km/s/Mpc},
\end{equation}
corresponding to a $3.3\%$ uncertainty.
So the $\chi^2$ of the Hubble constant data is
\begin{equation}
\chi^2_{h}=\left({h-0.738\over 0.024}\right)^2.
\end{equation}

\subsection{The total $\chi^2$}

Since the SNIa, CMB, BAO and $H_0$ are effectively independent measurements,
we can combine them by simply adding together the $\chi^2$ functions,
i.e.,
\begin{equation}
\chi^2_{All} = \chi^2_{SN} + \chi^2_{CMB} + \chi^2_{BAO} + \chi^2_{h}.
\end{equation}


\end{document}